\newcommand\beq{\begin{equation}}    
\newcommand\eeq{\end{equation}}    
\newcommand\bea{\begin{eqnarray}}    
\newcommand\eea{\end{eqnarray}}    
    
\newcommand\fettk{{\bf k}}   
\newcommand\fettn{{\bf n}}   
\newcommand\fettm{{\bf m}}   
\newcommand\fettu{{\bf u}}   
\newcommand\fettx{{\bf x}}   
\newcommand{\figinclude}[1]{\epsfverbosetrue\epsfxsize=\columnwidth\epsfbox{#1}}   
\documentstyle[aps,epsfig,multicol]{revtex}   
\begin{document}    
\draft    
\preprint{}   
\title{  
Transition to turbulence in a shear flow   
}    
    
\author{Bruno Eckhardt$^{1}$ and Alois Mersmann$^2$}    
\address{$^1$ Fachbereich Physik der Philipps Universt\"at   
Marburg, D-35032 Marburg, Germany}   
\address{$^2$ Fachbereich Physik und Institut f\"ur Chemie und Biologie des    
Meeres\\   
C.v.~Ossietzky Universit\"at, Postfach 25 03, D-26111 Oldenburg, Germany}    
    
\date{\today}    
\maketitle{ }  
\begin{abstract}    
We analyze the properties of a 19 dimensional Galerkin   
approximation to a parallel shear flow. The laminar flow with    
a sinusoidal shape is stable for all Reynolds numbers $Re$. For   
sufficiently large $Re$ additional stationary flows   
occur; they are all unstable. The lifetimes of finite amplitude    
perturbations show a fractal dependence on amplitude and Reynolds number.   
These findings are in accord with observations on plane Couette flow    
and suggest a universality of this transition szenario in shear flows.   
\end{abstract}    
     
\pacs{47.20.Ft, 47.20.Ky, 47.15.Fe, 05.45.+b}   
   
\begin{multicols}{2}  
\section{Introduction}   
In many flows the transition to turbulence proceeds via a sequence of    
bifurcations to flows of ever increasing spatial and temporal complexity.   
Analytical and experimental efforts in particular on layers    
of fluid heated from below \cite{Busse,Kosch} 
and fluids between rotating concentric cylinders \cite{Kosch,Swinney}  
have lead to the identification and verification of several routes to    
turbulence, which typically involve a transition from a structureless   
laminar state to a stationary spatially modulated one and then to more   
complicated states in secondary and higher bifurcations.    
   
Transitions in shear flows do not seem to follow this pattern  
\cite{Grossmann1996}. Typically,   
a transition to a turbulent state can be induced for sufficiently large   
Reynolds number with finite amplitude   
perturbations, just as in a subcritical bifurcation. However, in the   
most spectacular cases of plane Couette flow between parallel plates 
and Hagen-Poiseuille flow   
in a pipe \cite{DR}, there is no linear instability of the laminar profile    
for any finite Reynolds number that could give rise to a subcritical   
bifurcation. The turbulent state seems to be high dimensional   
immediately, without clear temporal or spatial patterns   
(unlike the rolls in Rayleigh-B\'enard flow). And the transition seems to   
depend sensitively on the initial conditions. Based on these characteristic   
features it has been argued that a novel kind of transition to    
turbulence different from the well-known three low-dimensional ones   
is at work \cite{GG1994}.   
   
Recent activity has focussed on three features of this transition:   
the non-normality of the linear eigenvalue problem 
\cite{GG1994,BB1988,TTRD1993,wal1,wal2,DM}, the occurence of   
new stationary states without instability of the linear profile 
\cite{Nagata1990,Nagata1997,CB1997,wal98} and   
the fractal properties of the lifetime landscape of perturbations   
as a function of amplitude and Reynolds number 
\cite{ES1997}. The non-normality of the   
linear stability problem implies that even in the absence of exponentially   
growing eigenstates perturbations can first grow in amplitude before   
decaying since the eigenvectors are not orthogonal.  
During the decay other perturbations   
could be amplified, giving rise to a noise sustained turbulence 
\cite{IF1993}.  
The amplification could also cause random fluctuations to grow to a size  
where the nonlinear terms can no longer be neglected \cite{wal2,DM}.  
Then the dynamics including the nonlinear terms could belong   
to a new asymptotic state, different from the laminar profile,   
perhaps a turbulent attractor. Presumably, this attractor 
would be built around stationary or periodic solution. 
Here, the observation of tertiary structures  
\cite{Nagata1990,Nagata1997,CB1997,wal98}   
comes in since they could form the basis for the turbulent state.   
Finally, the observation of fractality in the lifetime distribution    
suggests that the turbulent state is not an attractor but rather a repeller:   
Infinite lifetimes occur only along the stable manifolds of the    
repeller, all other initial conditions will eventually decay. Permanent   
turbulence would thus correspond to noise induced excitations onto   
a repeller.    
   
In plane Couette flow some of the features described above have been   
identified, but only with extensive numerical effort 
\cite{Nagata1990,Nagata1997,CB1997,ES1997}. The aim of the    
present work is to present a simple model that is based on the    
Navier-Stokes equation and captures the essential elements of the    
transition. It is motivated in part by the desire to obtain a   
numerically more accessible model which perhaps will provide   
as much insight into the transition as the Lorenz model  
\cite{Lorenz} for the   
case of fluids heated from below (presumably at the price of 
similar shortcomings).   
The two and three degree    
of freedom models proposed by various groups  
(and reviewed in \cite{BT1997}) to study the effects of    
non-normality mock some features of the    
Navier-Stokes equations considered essential by their inventors    
but they are not derived in some systematic way  
from the Navier Stokes equation. The model used here differs from 
the one proposed by Waleffe \cite{wal2} in the selection of 
modes. 
 
Attempts to built models for shear flows using Fourier modes immediately   
reveal an intrinsic difficulty: In the case of fluids heated from below   
the nonlinearity arises from the coupling of the temperature gradient   
to the flow field so that two wave vectors, ${\bf k}$ and  
$2{\bf k}$, suffice to obtain nonlinear couplings.  
In shear flows, the nonlinearity has to come from   
the coupling of the flow field with itself through the advection term   
$({\bf u}\cdot\nabla){\bf u}$. This imposes rather strong    
constraints on the wave vectors. At least three wave vectors   
satisfying the triangle relation ${\bf k}_1+{\bf k}_2+{\bf k}_3=0$   
are required to collect a contribution from the advection term.   
A minimal model thus has at least six complex variables. Three of these    
decay monotonically to zero, leaving three for a nontrivial dynamics.   
In the subspaces investigated (B.E., unpublished),  
the most complex behaviour found is   
a perturbed pitchfork bifurcation, which may be seen as a precursor   
of the observed dynamics: for Reynolds numbers below a critical value,   
there is only one stable state. Above that value a pair of stable and    
unstable states is born in a saddle-node bifurcation. The stable   
state can be excited through perturbations of sufficient amplitude.   
The basins of attraction of the two stable states are intermingled,   
but the boundaries are smooth.   
   
Thus more wave vectors are needed and they have to couple in a nontrivial   
manner to sustain permanent dynamics. The specific set of modes used is   
discussed in section \ref{sec_model}. It is motivated by boundary  
conditions for   
the laminar profile and the observation that wave vectors pointing   
to the vortices of hexagons satisfy the triangle conditions in a   
most symmetrical manner. Other than that the selected vectors are   
a matter of trial and error. In the end we arrive at a model    
with 19 real amplitudes, two force terms and 212 quadratic couplings.   
Without driving and damping the dynamics is energy conserving, as would be   
the corresponding Euler equation (suitably truncated). Moreover, the   
perturbation amplitudes can be put together to give complete flow fields.   
Thus the model has a somewhat larger number of degrees of freedom,   
but the dynamics should provide a realistic approximation to    
shear flows.   
   
The outline of the paper is as follows. In section  
\ref{sec_model} we present the    
model, in particular the selected wave vectors, the equations of motion   
and a discussion of symmetries. In section \ref{sec_dynamic}  
we focus on the dynamical   
properties of initial perturbations as a function of amplitude and   
Reynolds number. In section \ref{sec_stationary} we discuss  
the stationary states, their   
bifurcations and their stability properties. We conclude  
in section \ref{sec_conclusion} 
with a summary and a few final remarks.   
   
\section{The model shear flow}   
\label{sec_model} 
Ideal parallel shear flows have infinite lateral extension. Both in    
experiment and theory this cannot be realized. We therefore follow    
the numerical tradition and chose periodic boundary conditions in the    
flow and neutral direction. The flow is confined by parallel walls   
a distance $d$ apart. A convenient way to built a low dimensional model   
is to use a Galerkin approximation. Solid boundaries would require    
the vanishing of all velocity components and complicated Galerkin   
functions where all the couplings can only be calculated numerically.   
However, under the assumption that here as well as in many other situations   
the details of the boundary conditions effect the results only   
quantitatively but not qualitatively, we can adopt free-free   
boundary conditions on the walls and use simple trigonometric   
functions as basis for the Galerkin expansion. Similarly, the nature   
of the driving (pressure, boundary conditions or volume force) should   
not be essential so that we take a volume force proportional to   
some basis function (or a linear combination thereof). This still leaves   
plenty of free parameters to be fixed below.   
   
\subsection{Galerkin approximation}   
We expand the velocity field in Fourier modes,   
\beq   
\fettu(\fettx,t) = \sum_{\fettk} \fettu(\fettk,t) e^{i\fettk\cdot\fettx}\,.   
\eeq   
Incompressibility demands   
\beq   
\fettu(\fettk,t) \cdot \fettk = 0 \,.   
\label{incomp}   
\eeq   
The Navier-Stokes equation for the amplitudes $\fettu(\fettk,t)$ becomes   
\bea   
\partial_t\fettu(\fettk,t) = &-& i p_\fettk \fettk    
- i \sum_{{\bf p}+{\bf q}=\fettk} \left(\fettu({\bf p},t)\cdot {\bf q}   
\right) \fettu({\bf q},t) \nonumber\\   
&-& \nu \fettk^2 \fettu(\fettk,t) + f_\fettk  
\eea   
where $p_\fettk$ are the Fourier components of the pressure (divided by   
the density), $\nu$ is the kinematic viscosity and $f_\fettk$ are the    
Fourier components of the volume force sustaining the laminar profile.   
   
There are three constraints on the components $\fettu(\fettk)$:   
incompressibility (\ref{incomp}), reality of the velocity   
field,   
\beq   
\fettu(-\fettk) = \fettu(\fettk)^*   
\eeq   
and the boundary conditions that the flow is limited by two parallel,   
impenetrable plates. The ensuing requirement  
$u_z(x,y,z)=0$ at $z=0$ and $z=d$ (where $d$ is the separation between plates) 
is most easily implemented through   
periodicity in $z$ and the mirror symmetry   
\beq   
\left(\matrix{u_x \cr u_y \cr u_z}\right)   
(x,y,-z) =    
\left(\matrix{u_x \cr u_y \cr -u_z}\right)   
(x,y,z) \,,   
\label{mirrorsymm}   
\eeq   
which in Fourier space requires   
\beq   
\left(\matrix{u_x \cr u_y \cr u_z}\right)   
(-k_x,-k_y,k_z) =    
\left(\matrix{u_x^* \cr u_y^* \cr -u_z^*}\right)   
(k_x,k_y,k_z) \,.   
\eeq   
This is not sufficient to fix the coefficients: the dynamics also has to    
stay in the relevant subspace, and thus the time derivatives   
have to satisfy similar requirements.   
   
\subsection{The wave vectors}   
The choice of wave vectors is motivated by the geometry of the flow  
and the aim to include nonlinear couplings.   
The basic flow shall be a flow in $y$-direction, neutral  
in the $x$-direction and sheared in the $z$-direction.   
Thus we take the first three wave vectors in $z$-direction,   
\beq   
\fettk_1 = \left(\matrix{0\cr0\cr1}\right)\,, \qquad   
\fettk_2 = \left(\matrix{0\cr0\cr2}\right)\,, \qquad   
\fettk_3 = \left(\matrix{0\cr0\cr3}\right) \,.   
\eeq   
The negative vectors $-\fettk_i$ also belong to the   
set but will not be numbered explicitely.   
In these units, the periodicity in the $z$-direction is    
$2\pi$, so that the separation between the plates is $d=\pi$   
because of the mirror symmetry (\ref{mirrorsymm}).   
The amplitude $\fettu(\fettk_1)$ will carry the laminar profile and    
$\fettu(\fettk_3)$ can be excited as a modification to the laminar profile.   
$\fettk_2$ is needed to provide couplings through the nonlinear term.   
These three vectors satisfy a triangle identity   
$\fettk_1+\fettk_2-\fettk_3=0$, but the nonlinear term vanishes   
since they are parallel.   
   
The next set of wave vectors contains modulations in the flow and    
neutral direction,   
\beq   
\fettk_4 = \left(\matrix{1\cr0\cr0}\right)\,, \quad   
\fettk_5 = \left(\matrix{1/2\cr\sqrt{3}/2\cr0}\right)\,, \quad   
\fettk_6 = \left(\matrix{1/2\cr-\sqrt{3}/2\cr0}\right) \,.   
\eeq   
Together with $-\fettk_i$ they form a regular hexagon, so that they provide   
nontrivial couplings in the nonlinear term. The periodicity in     
flow direction is $4\pi/\sqrt{3}$, in the neutral direction it is   
$4\pi$.    
   
\begin{figure}    
   \figinclude{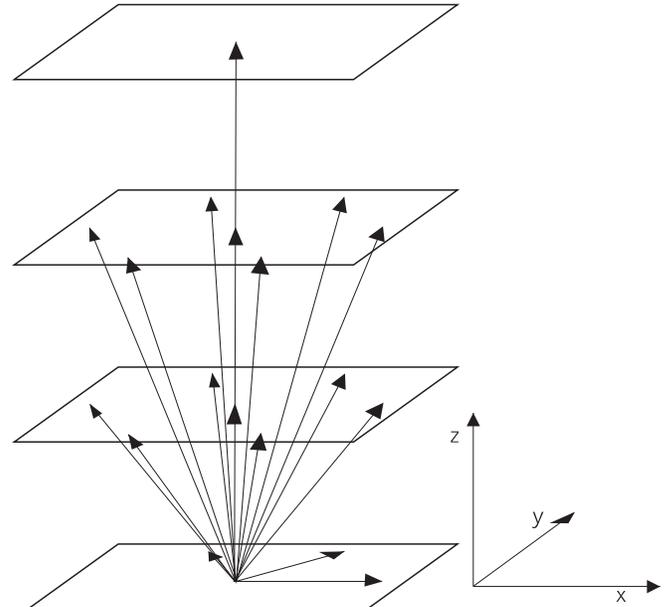}   
\caption[]{The 19 wave vectors $\fettk_1\ldots \fettk_{19}$. The full set   
is obtained by complementing with $-\fettk_i$. Thus, there are only three   
vectors on the symmetry plane $z=0$, six each on the two levels   
above and a single one on the third plane with $k_z=3$.   
}    
\label{wavevectors}    
\end{figure}

Finally, this hexagon is lifted upwards with $\fettk_1$   
and $\fettk_2$ to form the remaining 12 vectors,   
\bea   
\fettk_7\;\, &=& \fettk_1 + \fettk_4 \quad\;  
  \fettk_8\;\, = \fettk_1 + \fettk_5 \quad\;   
  \fettk_9\; = \fettk_1 + \fettk_6 \cr   
\fettk_{10} &=& \fettk_1 - \fettk_4 \quad\;   
  \fettk_{11} = \fettk_1 - \fettk_5 \quad\;   
  \fettk_{12} = \fettk_1 - \fettk_6 \cr   
\fettk_{13} &=& \fettk_2 + \fettk_4 \quad\;   
  \fettk_{14} = \fettk_2 + \fettk_5 \quad\;   
  \fettk_{15} = \fettk_2 + \fettk_6 \cr   
\fettk_{16} &=& \fettk_2 - \fettk_4 \quad\;   
  \fettk_{17} = \fettk_2 - \fettk_5 \quad\;   
  \fettk_{18} = \fettk_2 - \fettk_6\,.   
\eea   
The full set $\fettk_i$, $i=1\ldots18$ is shown in Fig.~\ref{wavevectors}.   
   
The Fourier amplitudes $\fettu(\fettk_i)$ have to be orthogonal to   
$\fettk_i$ because of incompressibility (\ref{incomp}). If they are   
expanded in basis vectors perpendicular to $\fettk_i$, the pressure   
drops out of the equations and need not be calculated. We therefore   
chose normalized basis vectors   
\bea   
\fettn(\fettk_i) &=& \left.\left({-k_x k_z\over k_x^2+k_y^2} ,    
  {-k_y k_z\over k_x^2+k_y^2}, 1\right)^T \right/  
  \sqrt{1+k_z^2/(k_x^2+k_y^2)} \cr    
\fettm(\fettk_i) &=& \left({k_y} ,    
  {-k_x}, 0\right)^T\left/ \sqrt{k_x^2+k_y^2} \right.\,  
\eea   
so that $\fettn$, $\fettm$ and $\fettk$ form an orthogonal set of
basis vectors.   
For the negative vectors $-\fettk_i$ we chose the basis vectors  
$\fettn(-\fettk_i)=\fettn(\fettk_i)$ and  
$\fettm(-\fettk_i)=-\fettm(\fettk_i)$.  
If the $x$ and $y$ components of $\fettk$ vanish, the above definitions   
are singular and replaced by   
\beq   
\fettn = (1,0,0)^T \qquad \fettm = (0,1,0)^T \,.   
\eeq   
The amplitudes of the velocity amplitude are now expanded as   
\beq   
\fettu(\fettk_i,t) = \alpha(\fettk_i,t) \fettn(\fettk_i) +    
\beta(\fettk_i,t) \fettm(\fettk_i) \,.   
\eeq   
   
The impenetrable plates impose further constraints on the    
$\alpha(\fettk_i)$ and $\beta(\fettk_i)$. For $i=1$, $2$ and $3$   
the wave vector has no components in the $x$- and $y$-directions,   
so that $\alpha$ and $\beta$ have to be real.   
For $i=4$, $5$ and $6$ the velocity field cannot have any components in the   
$z$-direction, hence $\alpha=0$.   
The remaining wave vectors $\fettk_i$ and $-\fettk_i$ with $i=7,\ldots,18$,   
a total of 24, divide up into six groups of 4 vectors each,   
\beq 
\fettk=(k_x,k_y,k_z),\; \fettk' = (-k_x,-k_y,k_z), \; -\fettk \;   
\mbox{and}\; -\fettk'\,.   
\eeq   
The groups are formed by the vectors and their negatives in the pairs   
with indices (7,10), (8,11), (9,12), (13,16), (14,17) and (15,18).   
The amplitudes of the vectors in the sets are related by   
\bea   
\alpha(\fettk) &=& \alpha(-\fettk)^* = - \alpha(\fettk')^* \cr   
\beta(\fettk) &=& \beta(-\fettk)^* = - \beta(\fettk')^* \,.   
\label{bcs}   
\eea   
Thus the full model has $6 + 6 + 6\times4 = 36$ real amplitudes.   
Restricting the flow by a point symmetry around 
$\fettx_0= (0,0,\pi/2)^T \,$ eliminates the contributions from  
${\bf k}_2$ and some other components, resulting in  
a 19-dimensional subspace with nontrivial dynamics  
and the following amplitudes:  
\bea   
\alpha(\fettk_1) &=& y_1 \qquad \beta(\fettk_1) = y_2\cr   
\alpha(\fettk_3) &=& y_3 \qquad \beta(\fettk_3) = y_4\cr   
\beta(\fettk_4) &=& i y_5 \qquad \beta(\fettk_5) = i y_6 \qquad    
\beta(\fettk_6) = i y_7 \cr   
\alpha(\fettk_7) &=& y_8 \qquad \beta(\fettk_7) = y_9\cr   
\alpha(\fettk_8) &=& y_{10} \qquad \beta(\fettk_8) = y_{11}\cr   
\alpha(\fettk_9) &=& y_{12} \qquad \beta(\fettk_9) = y_{13}\cr   
\alpha(\fettk_{13}) &=& i y_{14} \qquad \beta(\fettk_{13}) = i y_{15}\cr   
\alpha(\fettk_{14}) &=& i y_{16} \qquad \beta(\fettk_{14}) = i y_{17}\cr   
\alpha(\fettk_{15}) &=& i y_{18} \qquad \beta(\fettk_{15}) = i y_{19}\,;   
\eea  
components not listed vanish or are related to the given ones by the    
boundary conditions (\ref{bcs}).   
A complete listing of the flow fields
${\bf u}_i$ associated with the coefficients $y_i$  
such that ${\bf u}=\sum_i y_i {\bf u}_i$   
as well as of the equations of motion are available from the authors.
   
\subsection{The equations of motion}   
\begin{figure}    
   \figinclude{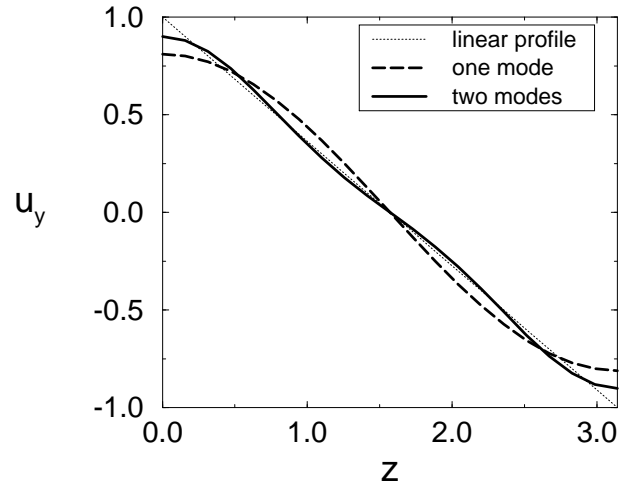}   
\caption[]{The laminar profile in case of one (a) or two driven modes (b),
compare eq.~\ref{laminar_profile}.
}    
\label{laminarflows}   
\end{figure}

In this 19-dimensional subspace $y_1\ldots y_{19}$ the equations of    
motion are of the form   
\beq   
\dot y_i = \sum_{j,k} A_{ijk} y_j y_k - \nu K_i y_i + f_i \,.   
\eeq   
Of the driving force all components but $f_2$ and $f_4$ vanish. 
Moreover, if the $f$'s are taken to be proportional to $\nu$, 
the resulting laminar profile has an amplitude independent of  
viscosity (and thus Reynolds number). 
These components give rise to a laminar profile that is a superposition 
of a $\cos(z)$ profile (from $f_2$) and a $\cos(3z)$ profile (from 
$f_4$).
This allows us to approximate the first two terms
of the Fourier expansion of a linear profile with velocity 
$u_y=\pm1$ at the walls, 
\beq   
\fettu_0= \frac{8}{\pi^2} (\cos z + {1\over 9} \cos 3z)    
\, {\bf e}_y \,. 
\label{laminar_profile} 
\eeq   
that can be obtained with a driving $f_2 = 4\nu/\pi^2$ and 
$f_4=4\nu/9\pi^2$ (see Fig.~\ref{laminarflows}).  
   
The nonlinear interactions in the Navier-Stokes equation conserve the   
energy $E=\frac{1}{2}\int dV \fettu^2$. In the 19-dimensional subspace, the   
corresponding quadratic form is   
\beq   
E = V \left( \sum_{i=1}^{7} y_i^2 + 2 \sum_{i=8}^{19} y_i^2\right)\,.   
\eeq   
The above equations conserve this form   
without driving and dissipation. With dissipation but still without 
driving, the time derivative is negative definite, indicating a  
monotonic decay of energy to zero.   
   
Finally, we define the Reynolds number using the wall velocity of  
the linear profile, $u_0=1$, the half width of the gap, 
$D=d/2=\pi/2$ and the viscosity $\nu$,  
\beq   
Re = u_0 D/\nu = \pi/2\nu \,.   
\eeq   
The other geometric parameters are a period $4\pi/\sqrt{3}$ in 
flow direction and $4\pi$ perpendicular to it. 
 
\subsection{Symmetries}  
\label{symmetries}  
We achieved the impenetrability of the plates   
by requiring the mirror symmetry:  
\beq   
\left(\matrix{u_x \cr u_y \cr u_z}\right)   
(x,y,-z) =    
\left(\matrix{u_x \cr u_y \cr -u_z}\right)   
(x,y,z) \, .  
\eeq   
  
The reduction from 36 to 19 modes was achieved by restricting the   
dynamics to a subspace where the flow has the point symmetry around  
$\fettx_0= (0,0,\pi/2)^T \,$, a point in the middle of the shear layer,  
\beq   
\left(\matrix{u_x \cr u_y \cr u_z}\right) (x,y,z+\pi/2)   
=\left(\matrix{-u_x \cr -u_y \cr -u_z}\right) (-x,-y,-z+\pi/2) \,.   
\eeq   
In addition, there are further symmetries that can be used to  
reduce the phase space.  
There is a reflection on the $y$-$z$-plane, 
\beq  
T_1:  
\left(\matrix{u_x \cr u_y \cr u_z}\right)   
(x,y,z) \rightarrow    
\left(\matrix{-u_x \cr u_y \cr u_z}\right)   
(-x,y,z) \,. \\  
\eeq  
and two shifts by half a lattice spacing,  
\bea  
T_2:& &  
\left(\matrix{u_x \cr u_y \cr u_z}\right)   
(x,y,z) \rightarrow    
\left(\matrix{u_x \cr u_y \cr u_z}\right)   
(x+2\pi,y,z) \, \\  
T_3:&&  
\left(\matrix{u_x \cr u_y \cr u_z}\right)   
(x,y,z) \rightarrow    
\left(\matrix{u_x \cr u_y \cr u_z}\right)   
(x+\pi,y+\pi/\sqrt 3,z) \, .  
\eea  
When applied to the flow these transformations induce changes 
in the variables $y_i$ (typically exchanges or sign changes), 
but the equations of motion are invariant under these transformations. 
Thus, if a certain flow has this symmetry, it leads to constraints 
on the variables $y_i$, and if it does not have this symmetry 
immediately a new flowfield can be obtained by applying 
this symmetry transformation. 
We do not attempt to analyze the full symmetry structure here and  
confine our discussion to two illustrative examples which are  
relevant for the stationary states discussed below. 
Demanding invariance of the flow field to the reflection symmetry $T_1$ 
leads to the following constraints on the variables $y_i$: 
\bea  
y_1&=&y_3=y_5=y_8=y_{15}=0\nonumber\\  
y_6&=&y_7\quad y_{10}=-y_{12} \nonumber\\  
y_{11}&=&y_{13}\quad y_{16}=-y_{18}\quad y_{17}=y_{19}.  
\eea  
The non vanishing components,  
$y_2$, $y_4$, $y_6=y_7$, $y_9$, $y_{10}=-y_{12}$,  
$y_{11}=y_{13}$, $y_{14}$,  
$y_{16}=-y_{18}$, $y_{17}=y_{19}$ thus define a 9 dimensional subspace. 
 
For the combined symmetry $T_1T_2$ we find the constraints 
\bea  
y_1&=&y_3=y_5=y_8=y_{15}=0\nonumber\\  
y_6&=&-y_7\quad y_{10}=y_{12} \nonumber\\  
y_{11}&=&-y_{13}\quad y_{16}=y_{18}\quad y_{17}=-y_{19}  
\eea  
and again a 9 dimensional subspace with non vanishing components  
$y_2$, $y_4$, $y_6=-y_7$, $y_9$, $y_{10}=y_{12}$, 
$y_{11}=-y_{13}$, $y_{14}$,  
$y_{16}=y_{18}$, $y_{17}=-y_{19}$. 
The dimensions of the invariant spaces vary from a minimum of  
6 (for each a $T_1T_3$ and $T_1T_2T_3$ invariance) and a maximum of  
10 (for $T_2T_3$-invariance). 
 
As mentioned, one can classify flows according to their symmetries. 
The most asymmetric flows are eightfold degenerate as the  
application of the eight combinations of the symmetries  
give eight distinct flows.  
The laminar flow profile is invariant under all  
the linear transformations and is the only member of   
the class with highest symmetry. The other stationary  
states discussed below fall into  
equivalence classes with eight members or four members if they  
are invariant under $T_1$ or $T_1T_2$.  
   
\section{Dynamics of perturbations} 
\label{sec_dynamic} 
 
\begin{figure}    
   \figinclude{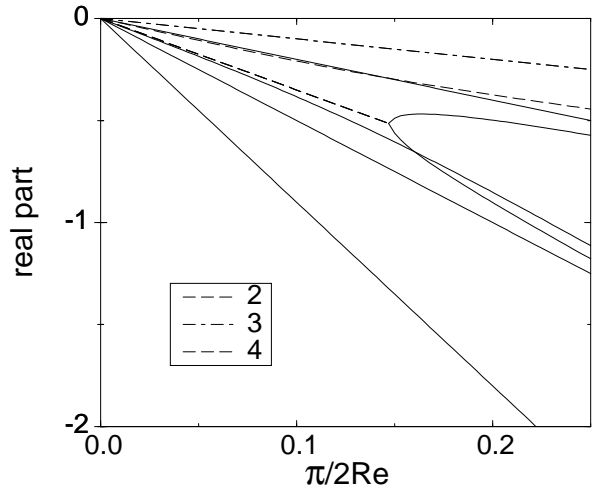}   
 \caption[]{Real parts of the eigenvalues of   
the linearized stability problem for one driven  
mode.
}    
\label{eigenvallaminar}   
\end{figure}

A stability analysis shows that the laminar flow profile is linearly  
stable for all Reynolds numbers. The matrix of the linearization is 
non-normal with a block structure along the diagonal. To bring this 
structure out more clearly, it is best to order the equations in the 
sequence 1, 2, 3, 4, 5, 7, 15, 8, 9, 14, 13, 19, 12, 18, 6, 11, 17, 
10, 16. 
The matrix of the linearization then is upper diagonal,
with a clear block structure: there are 10 eigenvalues 
isolated on the diagonal, three $2\times2$ blocks and one 
$3\times 3$ block as well as several couplings between them in the  
upper right corner. While some eigenvalues can be complex, all of them 
have negative real part as shown in  
Fig.~\ref{eigenvallaminar}. For vanishing viscosity, 
the eigenvalues become zero or purely imaginary. 
  
Large amplitude perturbations, however, need not decay. Already  
in the linear regime the non-orthogonality of the eigenvectors can  
give rise to intermediate amplifications into a regime  
where the nonlinear terms become important 
\cite{GG1994,BB1988,TTRD1993,wal1,wal2}. In a related  
study on plane Couette flow \cite{ES1997} 
we used the lifetime of perturbations  
to get information on the dynamics in a high-dimensional  
phase space. As in that case, the amplitude of the velocity  
field in the $z$-direction indicates the survival strength of  
a perturbation. Linearizing the equations of motion around the  
base flow ${\bf u}_0$ gives for the perturbation ${\bf u}'$  
the equation  
\beq  
\partial_t{\bf u}' = - ({\bf u}_0\cdot\nabla){\bf u}' -   
({\bf u}'\cdot\nabla){\bf u}_0- \nabla p'  
+ \nu \Delta {\bf u}' \,.  
\eeq  
The second term on the right hand side describes the energy source  
for the perturbation, and depends, because of  
${\bf u}_0 = u_0(z)\, {\bf e}_y$ and thus   
\beq  
({\bf u}'\cdot\nabla){\bf u}_0 = u_z' \partial_z u_0(z) {\bf e}_y  
\eeq  
in an essential way on the $z$-components of the perturbation.  
Thus, if the amplitudes $y_8$, $y_{10}$, $y_{12}$, $y_{14}$, $y_{16}$  
and $y_{18}$ become too small, the decay of the perturbation cannot  
be stopped any more. These modes account also for most of the 
off-diagonal block-couplings. A model for sustainable   
shear flow turbulence has to include some of these modes.  
  
\begin{figure}    
   \figinclude{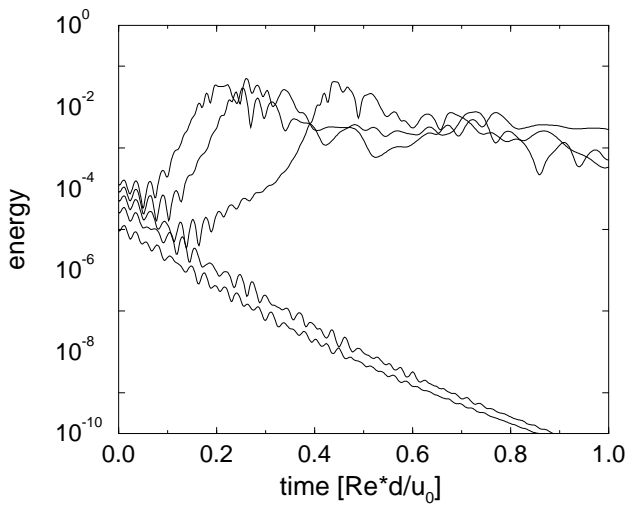}   
\caption[]{The dynamics of perturbations for one driven mode at $Re=400$.   
The perturbation was selected randomly and scale by factors
3, 5, 7, 9 and 11, from bottom to top.
}    
\label{timedependence}   
\end{figure}

We chose a fixed initial flow field with a random selection of  
amplitudes $y_1,\ldots,y_{19}$, scaled it by an amplitude parameter  
$A$ and measured the lifetime as a function of $A$ and  
Reynolds number Re. Fig.~\ref{timedependence}  
shows the time evolution of such a perturbation  
at $Re=400$ with one mode driven and for different amplitudes.  
For small $A$ there is an essentially exponential decay, whereas  
for  larger amplitudes the perturbation swings up to large  
amplitude  
and shows no sign of a decline at all. The results for  
many amplitudes and Reynolds numbers are collected in  
Fig.~\ref{landscape} in a landscape plot.  
For small Reynolds number and/or small amplitude  
the lifetimes of perturbations are short, indicated by the  
light areas. For Reynolds numbers around 100 isolated black spots  
appear, indicating the occurence of lifetimes larger than  
the integration time (which increases with $Re$ so that  
$t_{max}/Re = 4\pi$). The spottiness for $Re$ between about 100  
and 1000 is due to rapid changes in lifetimes from pixel to  
pixel. For Re above 1000 the long lifetimes dominate.  
These results are in good agreement with what has been observed in  
plane Couette flow. Fig.~\ref{landscape}b shows a similar plot for the case  
with two modes driven; it is qualitatively similar, but quantitatively  
shifted to higher Reynolds numbers.  
  
In connection with the non-normality of the linearized eigenvalue   
problem it has been argued that the upper limit on the    
size of perturbations for which the non-linear terms in the dynamics   
can be neglected decreases algebraically like $Re^{-\alpha}$. Different 
exponents have been proposed, ranging from 1 to 3  
\cite{GG1994,wal2,BT1997}. It seems that for large $Re$ (where the  
model is actually less reliable because of the limited spatial 
resolution) the envelope of the long lived states in the fractal 
life time plot decays like $Re^{-1}$. 
  
\begin{figure}       
\epsfysize=\columnwidth\epsfbox{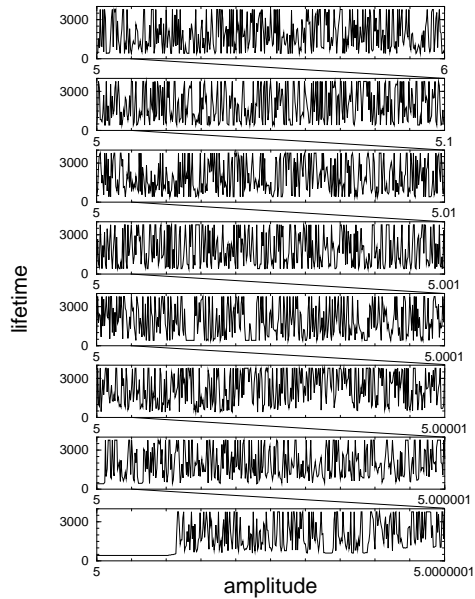}   
\caption[]{Lifetimes of perturbations as a function of amplitude for the case of 
one driven mode at $Re=200$ and successive magnifications by a factor of 10.
 }     
\label{lifetimes}   
\end{figure}

The sensitive dependence of lifetimes on initial conditions and  
parameters is further highlighted in  
Fig.~\ref{landscape_magnification} and \ref{lifetimes}.  
The first shows the lifetime in the plane of the amplitudes  
$y_{16}$ and $y_{17}$ at Reynolds number $Re=400$ with all  
other components fixed. There is considerable structure on  
many scales. One notes 'valleys' of short lifetimes between  
'plateaus' of longer lifetimes and a granular structure within  
both. The striations are reminiscent of features seen  
near fractal basin boundaries \cite{Ott}.  
Fig.~\ref{lifetimes} shows successive magnifications of lifetime  
versus amplitude plots at Re=200. Even after a magnification by  
$10^{7}$ there is no indication of a continuous and smooth variation  
of lifetime with amplitude.
  
\section{Stationary states}   
\label{sec_stationary} 
Motivated by the observation of new stationary structures 
in plane Couette flow for sufficiently high Reynolds number 
\cite{Nagata1990,Nagata1997,CB1997} we  
searched for  
non-trivial stationary solutions and studied their generation, evolution  
and symmetries.  
  
We computed the stationary states with the help of a Monte Carlo  
algorithm. The initial conditions for the $y_i$'s  
were chosen randomly out of the  
interval $[-1/2,1/2]$ and the Reynolds number was chosen randomly  
matrixwith an exponential bias for small $Re$ in the interval $[10,10000]$.  
With these initial conditions we entered a Newton algorithm.  
If the Newton algorithm converged, we followed the   
fixed point in Reynolds number as far as possible.   
We included about 200000 attempts in the Monte Carlo search.  
  
\begin{figure}    
   \figinclude{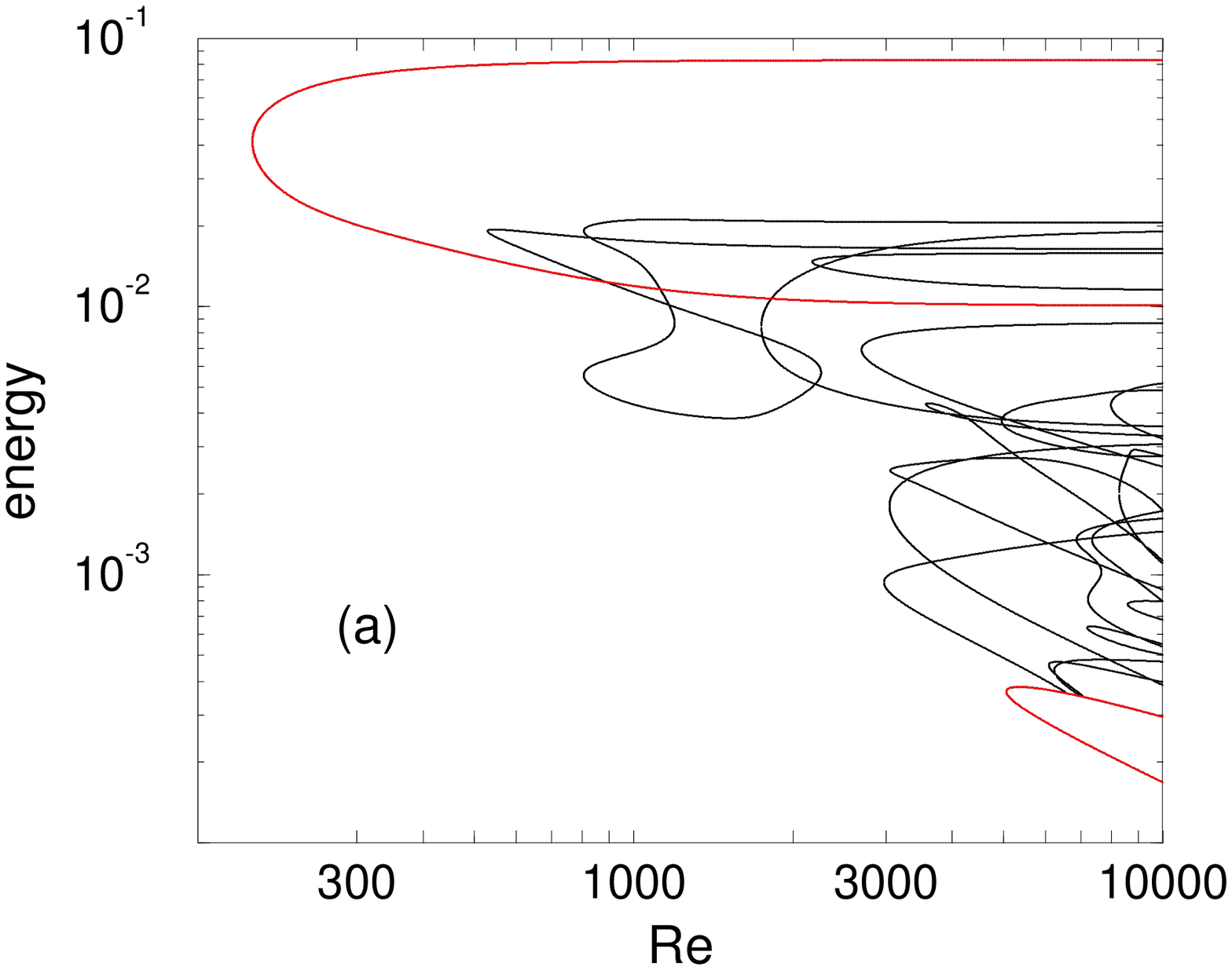}   
   \figinclude{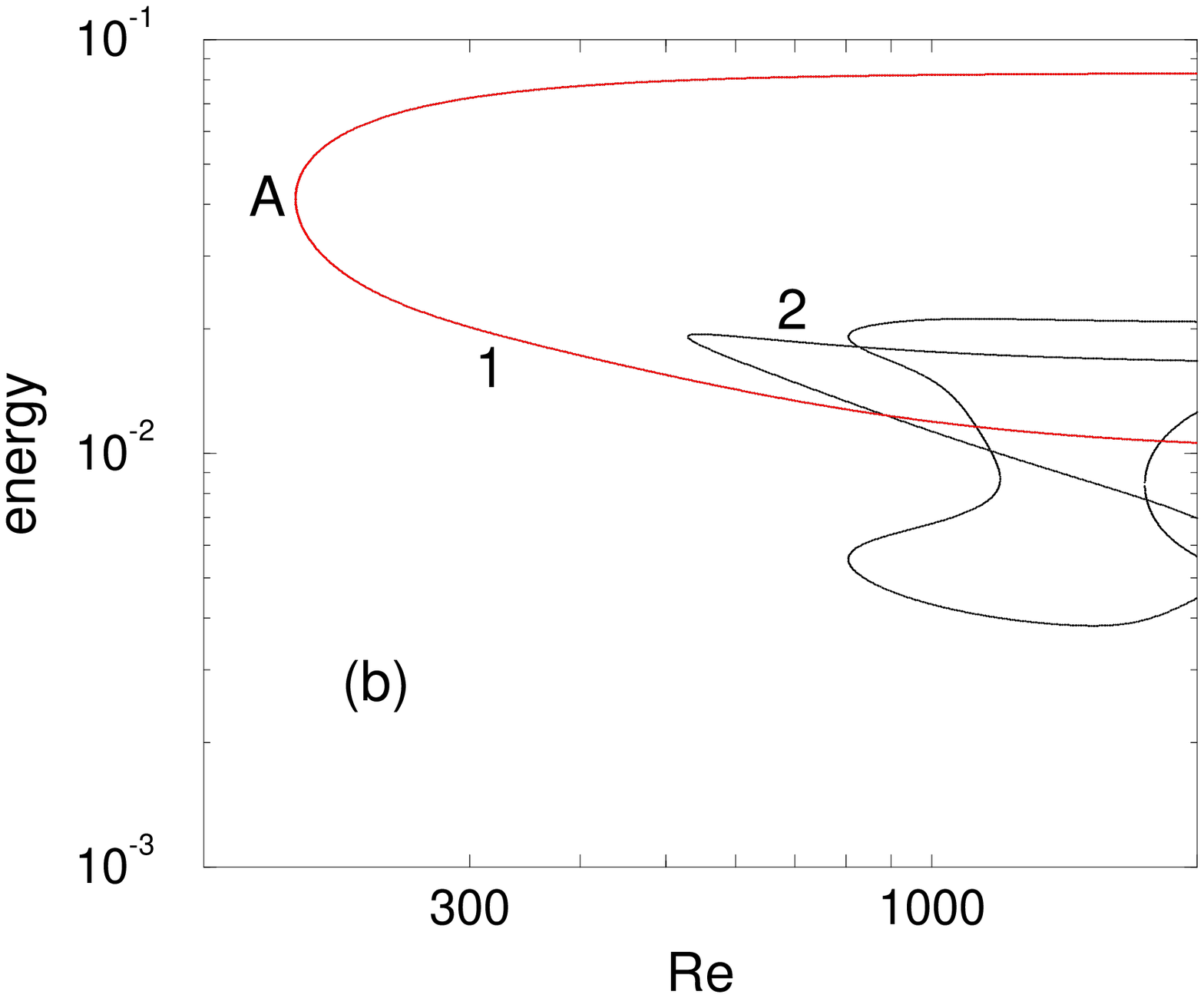}   
\caption[]{Stationary states for a single driven mode (a)  
and a magnification (b) near the leading saddle node
bifurcation near $Re=190$.   
}    
\label{stationary_single}   
\end{figure}

\begin{figure}    
   \figinclude{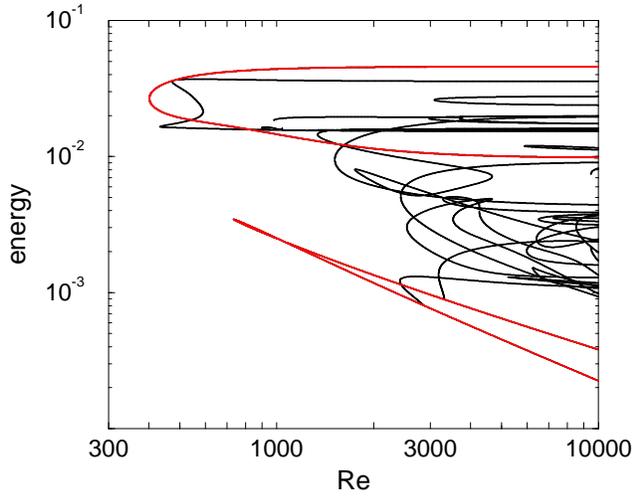}   
\caption[]{Stationary states for two driven modes. Compared to
Fig.~\ref{stationary_single} there seem to be more states and the 
next bifurcation is a lot closer to the leading one.
}    
\label{stationary_double}   
\end{figure}

The stationary states found for a single driven mode   
are collected in Fig.~\ref{stationary_single}. No  
stationary states (besides the laminar profile) were found for   
Reynolds numbers below about 190. Between 190 and about 500 there  
are eight stationary states which divide into two groups of  
four symmetry related states each. With increasing Reynolds number  
more and more stationary states are found and they reach  
down to smaller and smaller amplitude. The envelope of all states  
reflects the $Re^{-1}$ behaviour found for the borderline where  
nonlinearity becomes important. For two driven modes   
(Fig.~\ref{stationary_double}) the situation is similar.  
  
\begin{figure}    
   \figinclude{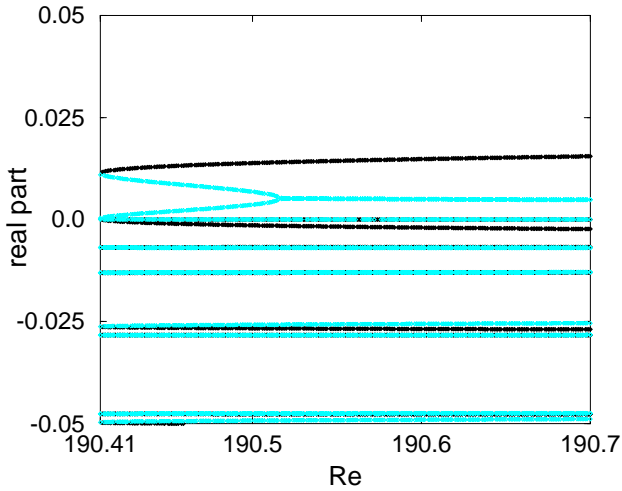}   
\caption[]{Eigenvalues of the two branches of the stationary states   
$a$ and $b$ near the saddle-node bifurcation around    
$Re\approx190$. Note that indeed two eigenvalues with real positive   
and negative real parts at zero, but there are also eigenvalues   
with positive real part.   
}    
\label{saddle_node_bif}   
\end{figure}

The appearance of the branches of the stationary states and   
in particular their coalescence near $Re=190$ suggests that the  
states are born out of a saddle-node bifurcation. And indeed,  
the eigenvalues as a function of $Re$ show two eigenvalues  
moving closer together and collapsing at zero for $Re=190.41$   
(Fig.~\ref{saddle_node_bif}).  
However, these eigenvalues are not the leading ones, so that  
one set of states has three unstable eigenvalues, the other  
two unstable ones. It is thus a `saddle-node' bifurcation into  
unstable states. 
  
\begin{figure}    
   \figinclude{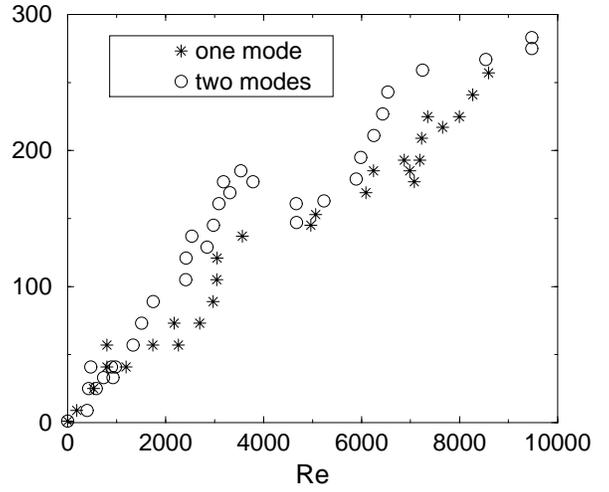}   
\caption[]{Proliferation of stationary states for one and two driven modes.   
}    
\label{proliferation}   
\end{figure}

With increasing $Re$ more and more stationary states appear,  
partly through secondary bifurcations, partly through additional 
saddle-node bifurcations. Their number increases rapidly with  
Reynolds number (Fig.~\ref{proliferation}) and this increase goes 
in parallel with the increase in density of long lived states,  
Fig.~\ref{landscape}. 
The detailed structure of the bifurcation diagram is rather  
complex and has not yet been fully explored. We note here 
that the various stationary states may be grouped according  
to their symmetries introduced in section \ref{symmetries}  
and that we found only stationary states which belong to    
equivalent classes with four or eight members.   
The stationary states of the classes with four  
members are invariant under  the 
transformation $T_1$ or $T_1T_2$.  
In addition, there are forward directed bifurcations   
generating two new branches with the   
same symmetry properties (eight or four member class) and inverse   
bifurcations of two branches belonging to eight member equivalent classes.  
We also found a backward directed bifurcation generating branches of   
an eight member class, which is born out of a four member class branch.   
The scenarios described above are marked   
in the bifurcation diagram Fig. \ref{stationary_single}.  
  
\section{Concluding remarks}  
\label{sec_conclusion}  
The few degrees of freedom shear model introduced here lies 
halfway between the simplest models of non-normality and  
full simulations. Its dynamics has turned out to be surprisingly rich.  
There are a multitude of bifurcations introducing new stationary states  
besides the laminar profile, there are secondary bifurcations, and the  
distribution of life times shows fractal structures on amazingly  
small scales. It seems that as one goes from the low-dimensional  
models\cite{TTRD1993,GG1994}  via the present one to 
full simulations one notes not only an increase in numerical 
complexity but also the appearence of  
qualitatively new features \cite{EMS1998}. 
 
The simplest models with very few degrees of freedom focus 
on the non-normality of the  
linearized Navier-Stokes problem and emphasize the amplification 
of small perturbations. If the non-linearity is included a  
transition to another kind of dynamics, sometimes as simple 
as relaxation to a stationary point, is found \cite{BT1997}. 
 
Next in complexity are models like the one presented 
here that share with the few degree of freedom models  
the amplification and the transition 
but the additional degrees of freedom allow for chaos.  
When nonlinearities become important the dynamics does not 
settle to a fixed point or a limit cycle but continues 
irregularly for an essentially unpredictable time.  
The time is unpredictable because of the fractal 
life time distribution which seems to persist down to  
amazingly small scale: tiny variations in Reynolds number  
or amplitudes of the perturbation can cause major variations in 
life times. This fractal behaviour is the new quality   
introduced by the additional degrees of freedom.  
Indications for this behaviour are seen in the  
experiments by Mullin on pipe flow \cite{Mullin}. 
It is interesting to ask just how few degrees of freedom are  
necessary to obtain this 
behaviour. Reducing our model to the $T_1$ subspace gives one with 
just 9 degrees of freedom (comparable in number and 
flow behaviour to the ones of Waleffe \cite{wal2}) that still 
shows this fractal life time distributions. Further reduction, 
as in the four mode model of \cite{wal2}, seems to eliminate them. 
 
The full, spatially extended shear flows share essential 
features with the model but add new problems. 
Spatially resolved simulations of the present model \cite{wal98} 
as well as plane Couette flow with rigid-rigid boundary 
conditions \cite{Nagata1990,CB1997} 
show the occurence of additional stationary states 
at sufficiently high Reynolds number that are unstable. 
A novel and as yet unexplained feature in spatially extended 
plane Couette flow, which we believe 
to be connected to the high dimensionality of phase space, 
is the difference between Reynolds numbers where the first 
stationary states are born (about 125 in units of half 
the gap width and half the velocity difference) and the  
ones where experiments begin to see long lived states 
(about 300--350) \cite{experiments}.  
 
The fractal life time distributions have obvious 
similiarities to chaotic scattering \cite{EA1988,E1988,Ott}. 
Drawing on this analogy one would like to identify permanent structures 
in phase space away from the laminar profile that could 
sustain turbulence. This has partly been achieved by 
the search for stationary states. Many have been found but 
irritatingly only for Reynolds numbers above about 190 while 
long lived states seem to appear much earlier. The solution 
to this puzzle must be periodic states and indeed we have found 
a few periodic states in a symmetry reduced model at lower 
Reynolds numbers, close to the  
occurence of the first long lived states. This suggests that  
the dynamical system picture that long lived states have 
to be connected to persistent structures in phase space  
is tenable. 
 
There are several features of the model that can be  
studied further. In particular, quantitative characterizations of  
the fractal life time distribution, visualizations of the flow 
field, a detailed analysis of the primary and secondary bifurcation, 
an investigation of the dependence on the aspect ratio of the  
periodicity cell are required and look promising. We expect the 
lessons to be learned from this simple model to be useful 
in understanding the dynamics of full plane Couette and other  
shear flows. Work along these directions continues. 
 
     
\narrowtext
   
%
   
%
   
%
   
%
   
\begin{figure}    
   \figinclude{regmit1-0-0-100.dat}  
   \figinclude{regmit1-1-0-100.dat}  
\caption[]{Lifetime of perturbations as a function of amplitude   
and Reynolds number for the case of one driven mode (a) and two  
driven modes (b).    
The black regions correspond to lifetimes   
larger than $T=4\pi \cdot Re$, the white regions to lifetimes   
shorter than $T/10$   
The grey levels interpolate linearly between these levels.   
}    
\label{landscape}   
\end{figure}    
%
   
\begin{figure}    
   \figinclude{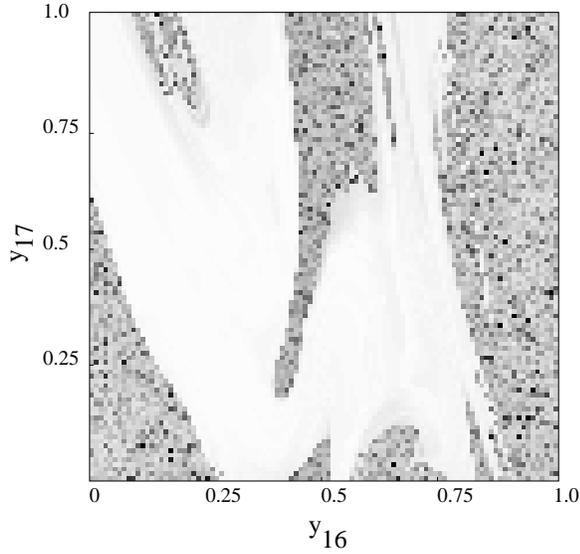}   
\caption[]{Magnification of the fractal landscape of lifetimes   
as a function of the amplitudes $y_{16}$ and $y_{17}$ for the 
same perturbation as in Fig.~\ref{landscape}
 at $Re=400$.   
}    
\label{landscape_magnification}   
\end{figure}    
%
   
%
   
%
   
%
   
%
   
\newpage  
  
\end{multicols}
\end{document}